# FINCH EYE: The Optical and Optomechanical Design of a GRISM-based SWIR Hyperspectral Imaging Payload for a 3U CubeSat

Iliya Shofman, Mario Ghio Neto, Theaswanth Ganesh, Kenya He, Aidan Armstrong, Ksenya Narkevich
University of Toronto Aerospace Team - Space Systems Division
55 St. George Street, Myhal Centre, Room 618, Toronto, Ontario, Canada
ss-outreach@utat.ca

## ABSTRACT

Crop residue is an important metric used for agricultural land-use monitoring and climate science research. Estimating crop residue coverage is essential to sustainable agricultural practices. There already exist scientific hyperspectral satellites, such as PRISMA and EnMAP, that can be used for crop residue estimation; however, they are very expensive to build and cannot be monopolized by individual farms. This leaves a gap in satellite remote sensing solutions which can be deployed operationally and at the farm scale. We seek to fill this gap by developing a low-cost satellite solution for crop residue estimation. The University of Toronto Aerospace Team is developing "FINCH EYE", the optical payload for the upcoming FINCH 3U CubeSat, to address this sensing gap.

We conceived of a novel ultra-compact push-broom architecture with a volume phase-holographic "grism" dispersive element to keep the design compact and simplify the mechanical assembly. The FINCH EYE will image hyperspectral data from 900nm to 1700nm at 10nm spectral resolution, with a spatial resolution of 100m, and a SNR of ~100. A preliminary feasibility analysis determined that these performance specifications can be achieved within the tight mass and volume constraints of a 3U CubeSat while potentially yielding scientifically valuable data.

In this paper, we will describe the optical design of FINCH EYE, which consists of a commercial objective lens, an InGaAs camera, and a custom lens-grism-lens spectrograph. We will also describe the optomechanical housing, emphasizing design features that facilitate proper alignment during assembly. Simulations in ANSYS ZEMAX OpticStudio verify the optical performance of the design we propose.

## 1. MOTIVATION FOR THE FINCH MISSION

The University of Toronto Aerospace Team (UTAT) Space Systems Division is currently developing a 3U CubeSat called FINCH (Field Imaging Nanosatellite for Crop residue Hyperspectral mapping). Our scientific mission is to provide a low-cost satellite solution for crop residue coverage estimation at the scale of entire farms. Crop residues play an important role in retaining soil moisture, preventing soil erosion, and can be indicators of tillage intensity.[1,2] Therefore, estimating crop residue coverage is essential for sustainable agriculture. On one hand, keeping crop residue to decompose after harvest can increase nutrients recycled into the soil.[3] On the other hand, excessive crop residue could harbor pathogens that may endanger subsequent harvests.[4] Hence, it is critical to quantify crop residue cover to inform agricultural decisions on the farm-scale.

A variety of crop residue estimation methods currently exist. Field surveying techniques can accurately measure crop residue cover, but are labor intensive, time consuming, and difficult to extrapolate to large areas of farmland. Aerial surveying methods with drones could cover larger areas but present difficulty in navigating terrain features, and the requirement of a licensed operator in some jurisdictions adds to recurring operating costs. Earth observation satellites can cover large geographic areas without the burden of manual or aerial surveying. Some scientific satellites, such as PRISMA and EnMAP, collect hyperspectral imagery that can be used for crop residue estimation; however, these satellites are very expensive to build and cannot be monopolized by individual farms. This leaves a gap in satellite remote sensing solutions which can be deployed operationally and at the farm scale.

CubeSats have emerged as a popular platform for scientific research, technology demonstrations, and Earth observation. The standardized form-factor facilitates ridesharing on rockets launching larger satellites, while the availability of commercial-off-the-shelf components have enabled substantially lower-cost and accelerated mission development lifecycles. The FINCH Mission seeks to provide an accessible solution for crop residue mapping for farmers by leveraging these recent technological advances.[5] The payload, FINCH EYE, seeks to demonstrate the technological and scientific feasibility of crop residue cover estimation on a low-cost nanosatellite form factor.

## 2. OPTICAL DESIGN OF FINCH EYE

The FINCH EYE is a compact push-broom hyperspectral imaging device built for a nanosatellite form-factor that is the first, to the author's best knowledge, to propose a volume-phase holographic (VPH) grism-based dispersive element. A "grism" is an optical device consisting of one or two prisms cemented onto the substrate of a diffraction grating. Compared to traditional ruled gratings, VPH gratings display high diffraction efficiencies for both polarizations, and are beneficial for spectroscopy applications where high brightness and good stray light rejection are necessary. The cemented prisms refract light in a direction opposite to the grating diffraction and hence maintain the overall direction of the optical axis. Ruled gratings in isolation would diffract light at some angle to the incoming beam and require fold mirrors or a bent optical layout, which are more challenging to implement on a volume-constrained nanosatellite platform. Hence, these characteristics make the VPH grism an appealing choice of the dispersive element for the nanosatellite hyperspectral Earth observation application.

### 2.1 Basic Principle of Operation

A schematic illustration of the proposed hyperspectral imager design is shown in Figure 1. Light from the Earth enters the aperture from the left and is imaged by the objective lens. A slit located in the objective's focal plane allows only one row of the Earth's image to be transmitted through the system, which is then re-imaged on the camera sensor by a one-to-one relay. A grism placed in the collimated space of the relay disperses white light in a direction orthogonal to the axis of the slit. Hence, spatial and spectral information would be projected onto the rows and columns of the camera sensor, respectively.

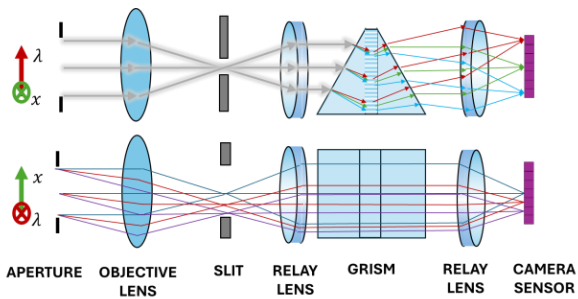

**Figure 1:** *diagram of proposed grism-based push-broom hyperspectral imager.*

Key performance specifications demanded from the FINCH EYE are summarized in Table 1. These preliminary values will be refined once a sensitivity analysis on our team's novel spectral unmixing algorithm for crop residue coverage estimation is complete.[6] While $\Delta\lambda$=10nm is a very coarse resolution, there are no sharp spectral features that must be resolved to adequately identify crop residue. Whereas farmland at the target imaging site is subdivided into half mile (~800m) square plots, the $\Delta x$=100m spatial resolution would yield ~20 usable pixels containing data entirely from within the farm field, given a 10ms sensor integration time setting and orbital inclination considered. While a finer spatial resolution is desired to resolve non-homogeneity within a farm field, this would require a longer focal length objective that would conflict with the limited payload volume allocation. The spectral range is determined by the detection range of the InGaAs camera sensor selected for this application. It should be noted that atmospheric absorption bands are present in this range. Signal-to-Noise Ratio (SNR) can be defined in many ways. We adopt the same definition as used in the EnMAP satellite, wherein the SNR is determined by the signal brightness given a constant and spatially uniform ground reflectance of $\rho$=0.3.[7] While typical minimum SNR values for hyperspectral Earth observation missions are around 100:1.[8] It is not currently known what minimum SNR is required by our crop residue coverage estimation algorithm.

**Table 1:** Optical system performance specifications for FINCH EYE.

| Specification | Value |
|---|---|
| Spatial Resolution ($\Delta x$) | 100m |
| Spatial Field-of-View | 10km |
| Spectral Resolution ($\Delta\lambda$) | 10nm |
| Spectral Range | 900nm-1700nm |
| Signal-to-Noise Ratio (SNR) | 100:1 |

Basic optical system definitions can thus be established from the above requirements. Given the relationship for ground sampling distance derived from geometric optics:

$$\theta_{\text{res}} = \frac{\Delta x_{\min}}{H} = \frac{p}{f} \qquad (1)$$

where $\Delta x_{\min}$ is the spatial resolution, $H$ is the altitude, $p$ is the pixel pitch, and $f$ is the system focal length. The angular resolution, $\theta_{\text{res}}$, must exceed the diffraction limited resolution $\theta_{\text{diff}}$, which in turn is determined by the diameter of the aperture D:

$$\theta_{\text{res}} \geq \theta_{\text{diff}} = 1.22\frac{\lambda}{D} \qquad (2)$$

where $\lambda$ is the optical wavelength. The ratio of focal length and aperture diameter determines the system f-number, or $F/\#$. Smaller values indicate a "fast" lens aspect ratio with a large diameter relative to the focal length. Fast lenses have finer resolution limits and yield higher image brightness given some sensor integration time. The system $F/\#$ determines the SNR through relations given in other literature.[9]

Given the Low Earth Orbit altitude of $H$=500km, the camera pixel pitch $p$=15μm, and the required ground sampling distance of $\Delta x_{min}$=100m, the system focal length should be 75mm. The required spatial field of view is 20mrad or ±0.6°. To yield an adequate SNR, the f-number should be about f/2.

## 2.2 Optical System Design

To reduce costs and development time, the optical design calls for a commercially available objective lens. Although a custom objective lens could relax some system requirements while meeting the requisite optical performance, its development is a substantial technical risk. The selection for long focal length (f>50mm) refracting SWIR objective lenses are very limited. Commercial long-focal length telescope objectives tend to have large diameters, while shorter-focal-length machine-vision lenses are aberration-corrected for a wider field of view than necessary for our application and therefore also bulky. While catadioptric lenses can offer very long focal lengths within reasonable physical lens dimensions, catadioptric lenses are more difficult to manufacture and have even more limited commercial availability in the SWIR range. After considering optical performance, mass, and physical dimensions, we have chosen the HYPSW10020640 lens from Hyperion Optics. This f=100mm f/2 lens has a longer focal length than originally called for, which allows for some margin on the spatial resolution. While commercially available off-the-shelf (COTS) optical components, such as objectives, have been demonstrated on similar nanosatellite missions, rigorous environmental testing is necessary to establish confidence in the satellite's operability once in orbit.[10]

The lens-grating-lens (LGL) relay follows the objective lens. Since the effective focal lengths of the two relay lenses are identical, the system's effective focal length is unchanged. The made-to-order grism will be manufactured by Wasatch Photonics; more details about the specifications and performance of the grism will be shared in a future publication. The relay optics will also be custom-made. The first relay lens must have excellent achromatic performance to properly collimate light transmitted through the slit. The second relay lens does not need the same chromatic aberration correction because the camera sensor may be tilted to compensate for the chromatic focal shift. Instead, the second relay lens should have better off-axis aberration correction as the range of ray angles would be larger after the grism. The LGL relay system was designed and optimized in ANSYS ZEMAX OpticStudio with five standard optical glass materials, all spherical surfaces, and one COTS singlet asphere lens.

A ray-trace diagram of the optical system is shown for on-axis spatial rays in Figure 2; the grism was rotated by 90° to show spectral dispersion in the illustration only. The objective lens was modelled as a paraxial lens; a contour of its physical shape is traced. The optical system is made more compact with two fold mirrors, such that the maximum length is 225mm. The on-axis spatial rays are dispersed by the grism so that different wavelengths are focused at different positions on the tilted camera sensor plane.

The spot diagram is shown in Figure 3, with the columns representing different wavelengths and the rows representing different spatial off-axis angles. The system was optimized at 0.92μm, 1.05μm, 1.25μm, and 1.60μm because these wavelengths correspond to the centers of atmospheric passbands in the SWIR range. While only ±0.6° spatial field of view is required, the optical system performs well even at larger angles.

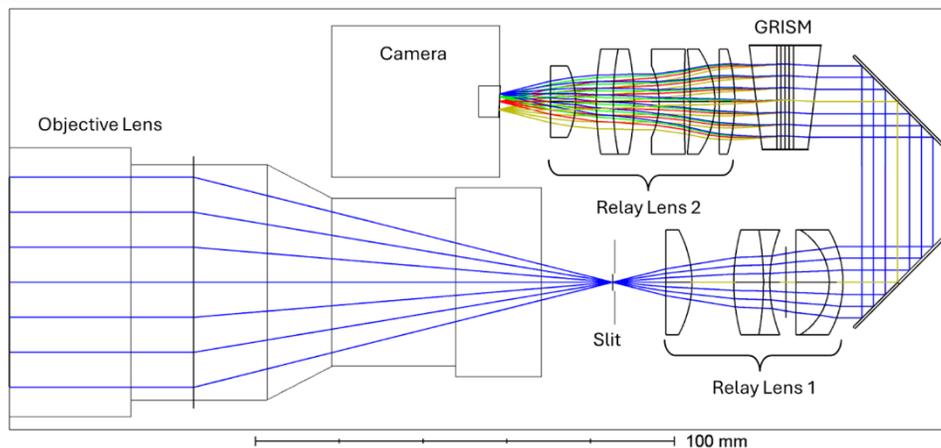

**Figure 2:** *Ray-trace diagram of FINCH EYE, with line traces showing the physical footprint.*

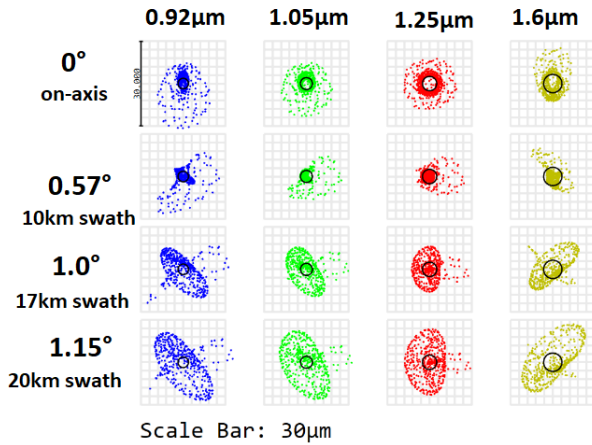

**Figure 3:** *Spot diagram at key wavelengths and several angles across the spatial field of view. The scale bar is twice the length of a single 15µm pixel.*

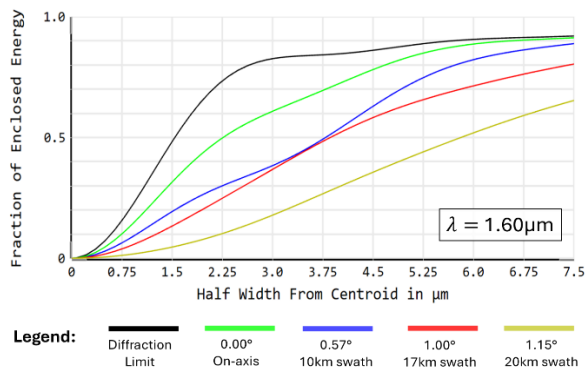

**Figure 4:** *Diffraction enclosed energy within a square pixel region for one wavelength, all incidence angles.*

Enclosed energy is another useful metric to assess imaging system performance and could be more informative than the RMS spot size alone. Enclosed energy indicates what proportion of energy carried by a light beam incident on the system at a specific angle and wavelength is enclosed within some area on the image plane. While a non-sequential ray tracing may be more accurate, Fast Fourier Transform (FFT)-based diffraction enclosed energy was used to quickly iterate during the design process. Figure 4 shows the diffraction "ensquared" energy, or the fraction of energy contained within a square region of a certain half-width. For a square pixel of 15µm pitch, the maximum half-width is 7.5µm before energy spills over into the neighboring pixel. The system displays nearly diffraction-limited performance for all wavelengths and the ±0.6° spatial field of view design requirements. Moreover, even at larger off-axis angles, the enclosed energy is still acceptable, with >80% energy enclosed within the pixel area.

The results for spot size and enclosed energy reflect only the nominal design. A tolerance analysis is being conducted to assess the impact of fabrication errors and mechanical misalignments. The nearly diffraction-limited performance is a good starting point for the design as it builds margin for performance degradation. Active alignment and compensation methods discussed in the optomechanical assembly may mitigate the impact of fabrication errors and mechanical misalignments.

## 3. OPTOMECHANICAL DESIGN OF FINCH EYE

The FINCH EYE optomechanical design combines commercial-off-the-shelf (COTS) components with custom-manufactured parts to achieve both passive and active alignment methods, utilizing miniaturized kinematic mounts and rotational adjusters. Figure 5 shows an overview of FINCH EYE, where the light first enters the system through the objective lens and propagates through the first lens barrel, reflecting on two mirrors situated on the optical platform, then enters the second lens barrel, finally exciting the sensor in the camera. This section explains the design intent behind each of these "modules" and outlines alignment mechanisms for each individual module in the assembly.

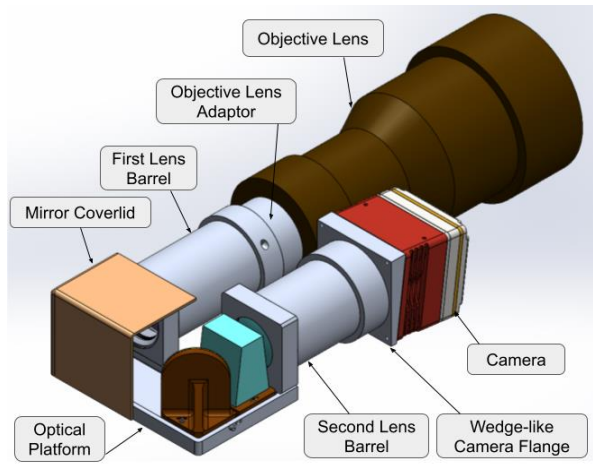

**Figure 5:** *Isometric view of the entire optomechanical assembly; mirror coverlid is shown in cross-section.*

### 3.1 Modular Assembly Design

It is important to design an optical system with a clear understanding of the subsequent steps of fabrication, assembly, and alignment. The proposed optical system can be conveniently split into several "modules" which can be assembled individually and connected with each other at mechanical interfaces. The two relay lenses will be assembled into individual lens barrels, while the fold mirrors and the grism will be mounted onto a planar platform. These individual lens barrel assemblies facilitate qualification testing of the relay lenses after manufacturing. The camera will be attached at an angle to the optical axis of the second relay lens using a wedge-like, flanged, threaded connector. Adding shims underneath the wedge and rotating the second relay lens within the mounting thread can assist with active alignment and re-focusing during assembly. Kinematics adjustments built into the fold mirror mounts could also compensate for tilt and decenter errors.

The two relay lenses will be assembled into individual lens barrels, with the first of these also housing the slit. Each lens barrel will be bored smooth and threaded at the ends where adjustment nuts are used. Each lens spacer will be machined to a close sliding fit with the inside of the barrel and will passively correct tilt and decenter errors created by the adjustment threads. [11]

The camera will be directly attached to a flange on the second lens barrel, utilizing its two alignment pins to ensure correct 'clocking', or rotation about the optical axis, with respect to the grism. This flange will be machined at an angle to the optical axis of the second relay lens barrel, similar to a wedge, to compensate for the different focal depths of each wavelength. Shim stock can be inserted between the camera and flange mating surfaces to fine-tune the angle of this wedge to a precision of 0.02°, utilizing a 24 μm shim.

Meanwhile, the fold mirrors and the grism will be mounted onto a planar platform, each on its removable kinematic mount. This allows each mirror to be aligned individually, utilizing attachment points for alignment instrumentation included in the platform design. The platform itself will be fabricated from stress-relieved material, as the presence of internal stresses can cause differential expansion during thermal cycling, resulting in undesired warping.

### 3.2 Alignment between Modules

This section outlines our alignment plan at critical points in the optical system. Adjustment is necessary because all manufactured parts, including COTS components such as lenses, are typically dimensioned to a tolerance that exceeds the acceptable optical tolerances, either directly or through tolerance stack-up. In other cases, acceptable mechanical tolerances can be achieved through costly manufacturing processes. The inclusion of an adjustable mechanism allows for a lower-precision, more cost-effective manufacturing process to be used instead, resulting in a lower total cost.

### 3.2.1 Slit-to-Relay Lens Axial Adjustment

The slit must be positioned in the focal plane of the first relay lens for the light to exit the lens properly collimated. Since the slit may be displaced axially during rotational adjustment, the first relay lens is allowed to translate axially within the lens barrel, as seen in Figure 6.

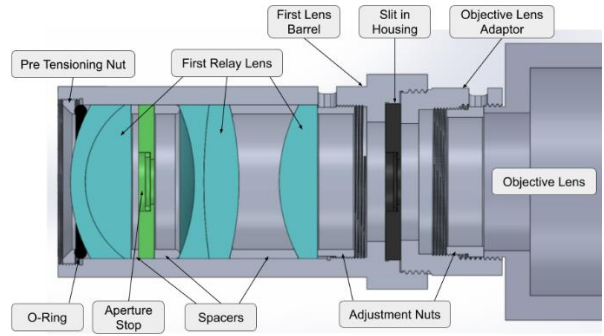

**Figure 6:** *Section cut of the first lens barrel. The objective lens is not shown in its entirety; it is modelled as a shell of the real product. Light propagates from the right to the left in this representation.*

At the leftmost end of the barrel, a preloaded O-ring is located between the relay lens and a pre-tensioning nut. This O-ring maintains axial compression of the lens group, absorbs differential expansion due to on-orbit thermal cycling between the glass and surrounding aluminum structures, and mitigates axial shock loads during launch. An adjustment nut at the rightmost end of the barrel mates with a 0.5mm pitch internal thread and contacts the leading singlet. When rotated, the adjustment nut either compresses or releases the lens stack, which is preloaded by the O-ring at the opposite end, ensuring the absence of backlash. Access to the adjustment nut is provided via a radial port in the barrel wall, allowing a tool to engage the outer surface of the nut. A millimeter scale etched around its circumference provides discrete alignment marks, supporting 152 rotational increments. This corresponds to a linear resolution of approximately 3.2 μm, given the 0.5 mm thread pitch, over a total travel range of ±1.5 mm. The light between the end of this lens group and the first fold mirror is collimated; therefore, the distance is not critical, and the mechanism can operate independently without cascading misalignments.

An identical mechanism is used inside the foreoptic adaptor, where an adjustment nut is used as a hard stop, setting the distance from the foreoptic to the slit with the same 3.2 μm precision. The same mechanism is implemented for the second lens barrel to correct the distance between the second relay lens and the camera sensor, thereby maintaining image focus.

### 3.2.2 Slit Rotational Adjustment

To avoid image distortion, the slit must remain aligned to the rows of camera sensor pixels to within 0.3°. However, pre-mounted slits suitable for the design do not have rotational alignment features. The COTS slit is mounted in a 3mm-thick, 1-inch-diameter aluminum housing, which will be modified to include a pair of orthogonal L-shaped cutouts that interface with an extra-fine pitch ball-end adjuster screw and a counterbalancing spring, as shown in Figure 7. This adjustment mechanism achieves precise rotation of 0.34° about the optical axis for every 90° turn of the adjuster screw, without generating significant radial forces.

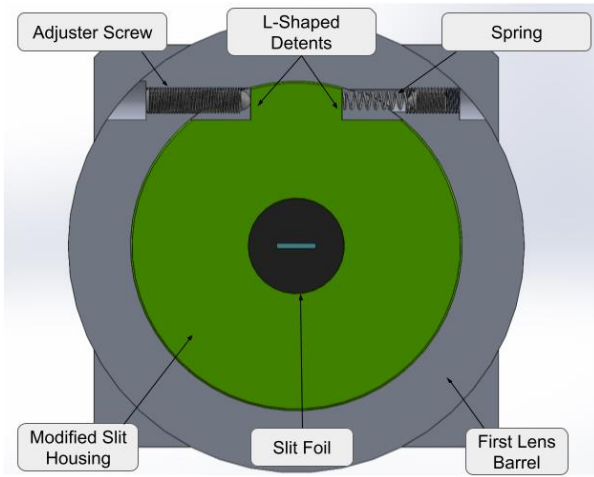

**Figure 7:** *Front view section cut of the slit adjuster mechanism. The slit has been made thicker to facilitate viewing in the figure.*

### 3.2.3 Fold Mirror and Grism Adjustment

Most commercial-off-the-shelf mirror mounts are prohibitively large and would protrude outside the allocated volume for FINCH EYE. This necessitates custom-designed mounts, seen in Figures 8 and 9, which allow a substantial reduction in part count and a 60% reduction in weight. Each 1-inch diameter fold mirror is glued to a kinematic mount integrated into the optical platform, capable of ± 2° of motion at a resolution of 0.5° per revolution, utilizing COTS M2x0.2 extra-fine-pitched adjustor screws. This combination permits correction for tilt and decenter at the exit of the fold mirror assembly.

Similarly, the grism is mounted on an integrated kinematic mount, allowing for tip and tilt adjustment with the same range and resolution as the mirror mounts. Rotation about the optical axis permits fine adjustment of the grism's 'clocking' with respect to the camera sensor, ensuring alignment of each spectral band with a row of pixels.

After assembly and alignment is complete, all moving parts will be fixed in place by applying low-outgassing adhesive to the adjustment threads. Thermally conductive compounds and conductive tape will be used to bridge gaps across the optomechanical housing, equalizing temperature gradients between the housing and the satellite chassis. [12]

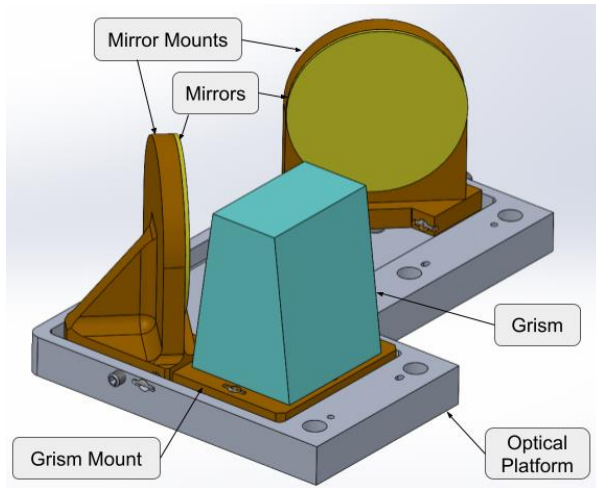

**Figure 8:** *Isometric view of the optical platform. Barrel mounting hardware and coverlid are not included here.*

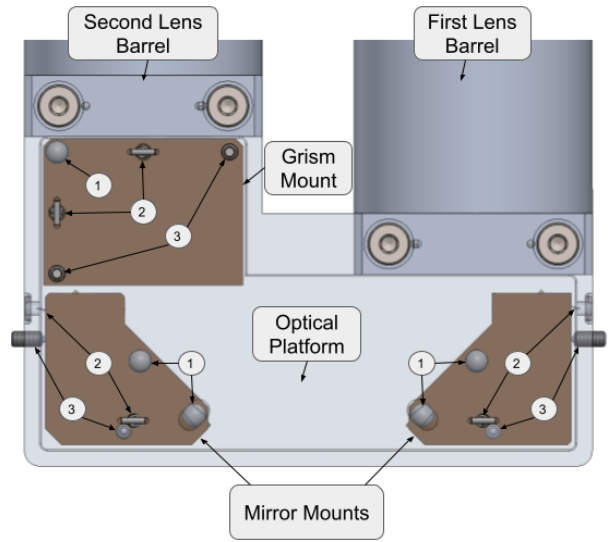

**Figure 9:** *Bottom view of the optical platform, mirror and grism mounts. The optical platform is translucent to facilitate viewing of the mechanisms in this figure: 1) ball bearings, 2) spring, 3) adjuster screw.*

### 3.3 Stray Light

Stray light mitigation features will be incorporated into the space between the two fold mirrors. A coverlid will be placed onto the platform onto which the mirrors are attached. This coverlid would be coated with an IR-absorbing coating, such as Vantablack™ or Acktar Vacuum Black™. Vanes for scattering stray light rays will be inserted into the free space that does not obstruct the intended light beam's clear aperture.

All cavities will be vented through simplified labyrinth vent ports, which mitigate the entrance of unwanted light by forcing at least three reflections until contact with any optical element. An example is shown in Figure 10, where light is forced to reflect, scatter, and be absorbed several times on the walls of the channels before entering the optical system. The impacts of these channels in airflow are easily calculated and were designed to make no appreciable impact on venting performance. [13]

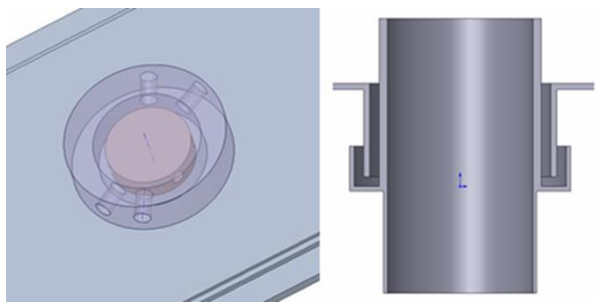

**Figure 10:** *Two types of labyrinth vent ports: a series of concentric channels (left), a channel with a series of 90° angles (right).* [13]

## 4. CONCLUSION

The FINCH EYE is designed to address a gap in farm-scale Earth observation: the need for a low-cost, remote sensing solution for crop residue mapping. While a variety of estimation methods currently exist, none are both affordable and convenient at the individual farm scale. This scientific objective could be met with the technical requirements and proposed design of FINCH EYE discussed herein.

In this paper, we introduced a compact push-broom SWIR hyperspectral instrument for a 3U CubeSat. The optical system employs a commercial objective lens and a custom volume-phase holographic grism-based spectrograph, making FINCH EYE a first-of-its-kind design for hyperspectral Earth-observation nanosatellites. We also developed a modular optomechanical housing that is easy to fabricate, assemble, and align. The alignment features, such as an in-plane rotation module for the slit, precise axial adjustment within lens barrels, and custom kinematic mounts for fold mirrors, allow us to choose standard machining tolerances while being able to fine-tune optical performance. We incorporate stray light reduction solutions with an IR-absorbing coated coverlid, scattering vanes, and labyrinth venting holes.

We are currently in the process of refining our optical and optomechanical designs. We seek to increase the spatial field of view and spectral resolution of the instrument to make it more versatile for other scientific applications. Meanwhile, we are also engaging with custom-optical fabrication companies and machine shops to make our design more manufacturable. Our next step is to conduct an optical tolerance analysis, structural failure analyses, and a thermal analysis to validate the performance of our payload in the expected outer-space environment. Our findings will inform important manufacturing aspects such as material choices and mechanical tolerances for the optomechanical housing. Meanwhile, we are building a simple and low-cost volume-unconstrained line-scan hyperspectral imager for visible wavelengths to validate our design process and practice assembly, alignment, and testing.

## 5. ACKNOWLEDGMENTS

We wish to thank the University of Toronto Engineering Society (EngSoc) for their continued recognition and financial support, and University of Toronto Student Union (UTSU) for supporting the operations of UTAT Space Systems through the UTAT Innovation Fund.

We gratefully acknowledge our technical, legal, and administrative advisors for their generous contributions of time and expertise, which have been critical in addressing key challenges and advancing our mission.

Finally, we extend our most sincere thanks to all the members of UTAT Space Systems. Their dedication and commitment to the team has fostered a supportive and innovative environment that drives the continued development of FINCH.